\documentclass[aps,pra,reprint,fleqn,amsmath]{revtex4-2}

\usepackage{graphicx}
\usepackage[charter]{mathdesign}
\DeclareMathAlphabet{\pazocal}{OMS}{zplm}{m}{n}
\def\cal#1{\pazocal{#1}}
\def\d{{\mathrm d}}
\def\tint{\mbox{\large$\int$}}
\def\cbone{\mbox{${\cal C}_{1}$}}
\def\cbtwo{\mbox{${\cal C}_{2}$}}
\def\vbone{\mbox{${\cal V}_{1}$}}
\def\vbtwo{\mbox{${\cal V}_{2}$}}

\def\eps{\epsilon}
\def\phit{\psi}

\begin{document}

\title{Enhanced high-order harmonics through periodicity breaks:\\
from backscattering to impurity states}
\author{Chuan Yu}
\author{Ulf Saalmann}
\author{Jan M. Rost}
\affiliation{Max-Planck-Institut f{\"u}r Physik komplexer Systeme, N{\"o}thnitzer Str.\ 38, 01187 Dresden, Germany}
\date{\today}

\begin{abstract}\noindent
Backscattering of delocalized electrons has been recently established [Phys.\,Rev.\,A 105, L041101 (2022)] as a mechanism to enhance high-order harmonic generation (HHG) in periodic systems with broken translational symmetry. Here we study this effect for a variable spatial gap in an atomic chain. Propagating the many-electron dynamics numerically, we find enhanced HHG and identify its origin in two mechanisms, depending on the gap size, either backscattering or enhanced tunneling from an impurity state. Since the gapped atomic chain exhibits both impurities and vacancies in a unified setting, it provides insight how periodicity breaks influence HHG in different scenarios.
\end{abstract}

\maketitle

\section{Introduction}\label{sec:intro}
\noindent
Since the first experimental demonstration of high harmonic generation (HHG) from a bulk crystal \cite{ghdi+11}, HHG in solid-state systems has attracted considerable interest (see, e.g., reviews \cite{vabr17,ghre19}), opening new possibilities to generate coherent extreme ultraviolet (XUV) light \cite{hamo+17,gaki+18} and to probe ultrafast dynamics in the target systems \cite{luga+15,hola+15,biha+21}.
HHG in a condensed-matter environment shares similarities with its well-known counterpart in atoms and molecules, but also differences and richer physics due to the larger structural complexity and variability \cite{ndgh+16,baha18,siji+19}. For example, the solid-state HHG cutoff was found to scale quasi-linearly with the driving field strength rather than quadratically as in atoms and molecules \cite{ghdi+11,vamc+15,wugh+15,naci+19}; under suitable conditions, solid-state HHG spectra exhibit multiple plateaus due to the multiple conduction bands \cite{ndgh+16,iksh+17,hade+17,yuir+20}.
The physical mechanism underlying HHG from solids, although still being lively discussed from different perspectives \cite{vamc+14,hist+14,tais+16,tamu+17,osch+17,chik20,laki+20,lila+21}, is typically described within a band picture in terms of intra- and interband processes \cite{vamc+14,vabr17,ghre19} acknowledging the delocalized nature of the participating electrons.
It has been demonstrated that a key concept in strong-field laser physics, the three-step model \cite{co93,leba+94}, can be adapted to describe laser-solid interactions with the band structures taken into account \cite{vamc+14,vamc+15} which explains the similarities to gas-phase HHG.

Several studies have utilized specific solid-state features to increase the HHG efficiency, aided by the progress in generating ultrashort XUV pulses and extracting ultrafast dynamics in condensed matter.
For example, experimental techniques utilizing nanostructures were reported to be capable of enhancing the HHG in semiconductors \cite{vagh+17,sita+17}.
HHG enhancement can be achieved also by introducing suitable dopants that increase the tunneling probability into conduction bands \cite{yuha+19,nefr+21}, or by using two-color schemes for the driving laser fields \cite{lizh+17,nath20,brna+21}.
Recently, we have demonstrated backscattering of delocalized electrons from system edges as a new mechanism to extend the HHG cutoff \cite{yusa+21},
which brings the mechanism of producing high-energy electrons in above-threshold ionization of atomic systems \cite{pabe+94,bego+18,waca+20} into the solid-state HHG context based on a band picture of delocalized electrons.
The backscattering mechanism should in general apply to quasi-periodic systems, and therefore we expect that many other sources of breaking the translational symmetry, such as impurities, domain walls or grain boundaries, will also induce backscattering-type HHG.

In this work we seek a general and systematic understanding how periodicity breaks, induced by different physical circumstances, enhance high harmonics through backscattering of delocalized electrons.
To this end, we extend the one-dimensional chain of atoms \cite{yusa+21} to mimic donor-type impurities \cite{yuha+19} and vacancy defects \cite{irha+20}.
We will show that an internal boundary caused by the periodicity break, which effectively divides the entire system into subsystems,
can indeed contribute to the high harmonics through backscattering of delocalized electrons similarly to the edge of the system.
This requires the subsystem sizes to be suitable for the backscattering mechanism, and it is desirable to have vacancy-like gaps between the subsystems in order to influence the backscattering effect efficiently.
Alternatively, if the periodicity is broken by a donor-type impurity, which makes the highest occupied orbital an impurity state with isolated energy between the valence and conduction bands, an overall enhancement of the HHG can be achieved.
This enhancement typically originates from the contribution of the highest-occupied impurity state, since the smaller energy gap between the conduction band and the impurity state leads to a much larger tunneling probability.
With the model introduced in this work, we will reveal how the HHG response changes through a continuous variation of the impurity-state energy, which therefore provides more insights into the impurity-induced HHG enhancement demonstrated in previous work \cite{yuha+19}.

The article is organized as follows.
In Sec.\,\ref{sec:theory}, we outline the theoretical approach and describe in particular how the internal boundary is introduced with a single parameter.
In Sec.\,\ref{sec:recap}, we give a brief description of the backscattering-type HHG demonstrated in our previous work \cite{yusa+21}.
Then the main results of this work are presented and discussed in Sec.\,\ref{sec:result}, namely how the
evolution from impurity to backscattering dominated HHG dynamics with increasing internal gap changes the corresponding HHG spectra.
Finally, Sec.\,\ref{sec:conclu} summarizes the conclusions.
Atomic units (a.u.) are used throughout unless otherwise indicated.

\section{Theoretical realization}\label{sec:theory}
\noindent
We consider $N$ nuclei with charge $Z$ located at positions $\{{x}_{i}\}$. The corresponding ionic potential reads
\begin{equation}\label{eq:ionpot}
v_{\text{ion}}^{}(x) = -\sum_{i=1}^{N} \frac{Z}{\sqrt{(x-{x}_{i})^{2}+\eps}},
\end{equation}
with a soft-core parameter $\eps$ to avoid singularities in the 1D treatment.

In Kohn-Sham (KS) DFT \cite{ul12}, the field-free electronic state is described by a set of KS orbitals fulfilling
\begin{subequations}
\begin{equation}\label{eq:stat_kseq}
\left\{-\frac{1}{2}\frac{\partial^{2}}{\partial\!x^{2}} + v_{\text{KS}}^{}[n](x)\right\}\varphi_{j}^{}(x) = \varepsilon_{j}^{}\varphi_{j}^{}(x),
\end{equation}
with the static KS potential
\begin{equation}\label{eq:stat_kspot}
v_{\text{KS}}^{}[n](x) = v_{\text{ion}}^{}(x) + v_{\text{H}}^{}[n](x) + v_{\text{xc}}^{}[n](x).
\end{equation}
\end{subequations}
Note that we restrict ourselves to charge and spin neutral systems, and therefore use the spin-restricted scheme for brevity.
With $N_{\text{occ}}\,{=}\,NZ/2$ denoting the number of occupied spatial orbitals, the total density is $n(x)\,{=}\,2\sum_{j=1}^{N_{\text{occ}}}|\varphi_{j}^{}(x)|^{2}$, where the factor of 2 accounts for the spin degeneracy.
The Hartree potential reads
\begin{equation}\label{eq:harpot}
v_{\text{H}}^{}[n](x) = \tint\!\d{x'} \frac{n(x')}{\sqrt{(x-x')^{2}+\eps}},
\end{equation}
and the exchange-correlation potential is treated in local density approximation (LDA)
\begin{equation}\label{eq:xcpot}
v_{\text{xc}}^{}[n](x) \simeq v_{\text{x}}^{}[n](x) = -[3n(x)/\pi]^{1/3}.
\end{equation}
Hence, our model is self-consistently constructed once the parameters $Z$, $\eps$, and $\{{x}_{i}\}$ are specified.
A typical configuration for this model 
\cite{hade+17,haba+18,yuha+19,wafe+19,lizh+19,libi20,irha+20,yuir+20,jema21,jeir+21} is a chain with parameters $Z\,{=}\,4$, $\eps\,{=}\,2.25$, and $\{{x}_{i}\}$ equally spaced by a lattice constant $d\,{=}\,7$.
In this work we consider this regular chain configuration as a prototype, and introduce an additional variable gap $\delta$ in the center of the chain.
This leads to a system composed of two subchains, which we refer to in the following as ``double-chain''.
We assume the two subchains to have the same configuration, with the total number of atoms $N$ chosen to be even.
The ionic positions for the double-chain read
\begin{equation}\label{eq:ionloc}
{x}_{i} = 
\begin{cases}
[i-\frac{N+1}{2}]d-\delta/2, & i=1,\cdots,\frac{N}{2}\\
[i-\frac{N+1}{2}]d+\delta/2, & i=\frac{N}{2}{+}1,\cdots,N
\end{cases}.
\end{equation}
Note that $\delta\,{=}\,0$ corresponds to the special case of a regular chain of $N$ equidistant atoms (referred to as a ``single chain'' or ``gapless chain''), while $\delta\,{\ne}\,0$ creates an internal boundary that breaks the periodicity.
We consider $\delta$ values in the range of $\delta\,{\geq}\,{-}d$, where the smallest value $\delta= -d$ means that the two atoms with indices $i\,{=}\,\tfrac{N}{2}$ and $i\,{=}\,\tfrac{N}{2}{+}1$ are on top of each other at $x\,{=}\,0$.

We let the double-chain interact with a laser pulse linearly polarized along the $x$-axis,
which is described by a vector potential within the dipole approximation,
\begin{equation}\label{eq:vecpot}
A(t) = A_{0}\sin^{2}\left(\frac{\omega_{0}t}{2 n_{\text{cyc}}}\right)\sin(\omega_{0}t)
\end{equation}
for $0 \leq t \leq {2\pi n_{\text{cyc}}}/{\omega_{0}}$ and $A(t)=0$ otherwise.
Hereby, $\omega_{0}$ is the angular frequency (photon energy) and $n_{\text{cyc}}$ is the number of cycles.
According to the TDDFT \cite{ul12}, the laser-driven system is governed by the time-dependent KS equations
\begin{subequations}\label{eq:dyn_ks}
\begin{align}\label{eq:dyn_kseq}
& {\mathrm i\,}\frac{\partial}{\partial\!t}\phit_{\!j}^{}(x,t) 
\notag \\ 
& = \left\{-\frac{1}{2}\frac{\partial^{2}}{\partial\!x^{2}} -{\mathrm i\,}A(t)\frac{\partial}{\partial\!x} + \widetilde{v}_{\text{KS}}^{}[n](x,t)\right\}\phit_{\!j}^{}(x,t),
\end{align}
where the KS potential,
\begin{equation}\label{eq:dyn_kspot}
\widetilde{v}_{\text{KS}}^{}[n](x,t) = v_{\text{ion}}^{}(x) + v_{\text{H}}^{}[n](x,t) + v_{\text{xc}}^{}[n](x,t),
\end{equation}
\end{subequations}
is determined by the time-dependent density $n(x,t) = 2\sum_{j=1}^{N_{\text{occ}}}|\phit_{\!j}^{}(x,t)|^{2}$.
As in previous works \cite{hade+17,yuir+20,yusa+21}, our considered laser interactions typically do not cause significant changes to the density;
therefore we simply assume that $\widetilde{v}_{\text{KS}}^{}[n](x,t)$ in Eq.\,\eqref{eq:dyn_kspot} remains the same as the initial $v_{\text{KS}}^{}[n](x)$ in Eq.\,\eqref{eq:stat_kspot}.
This frozen-KS approach has been found to capture basically the same HHG features as the dynamic-KS approach \cite{hade+17}, and it implies an independent-electron picture which has been frequently assumed in many studies on HHG in solids based on TDSEs \cite{iksh+17,lifu+18,jili+19,naci+19,nath20,wabi21}.
Details of the numerical methods for solving Eq.\,\eqref{eq:dyn_ks} can be found elsewhere \cite{yuir+20}.

For the calculation of the HHG spectra, we compute the total time-dependent current
\begin{equation}\label{eq:cur_tot}
J_{\text{tot}}^{}(t) = 2\sum_{j=1}^{N_{\text{occ}}}\!\tint\!\d{x}\,\text{Re}\Big[\phit_{\!j}^{*}(x,t)\Big(-{\mathrm i}\frac{\partial}{\partial\!x}+A(t)\Big)\phit_{\!j}^{}(x,t)\Big].
\end{equation}
The HHG spectral intensity is then evaluated as the modulus square of the Fourier-transformed current,
\begin{equation}\label{eq:hhg_tot}
S_{\text{tot}}^{}(\omega)\propto \left|\tint\d{t}\,W(t)J_{\text{tot}}(t)\exp(-{\mathrm i}{\omega}t)\right|^{2},
\end{equation}
where $W(t)$ is a window function of the pulse-envelope shape introduced to improve the signal-to-noise ratio.
When comparing the HHG spectra in systems with different number of atoms, it is natural to compute the HHG per atom \cite{yuir+20,yusa+21},
\begin{equation}\label{eq:hhg_atom}
S_{N}(\omega) = N^{-2}S_{\text{tot}}^{}(\omega).
\end{equation}

\begin{figure*}[t]
\includegraphics[width=\textwidth]{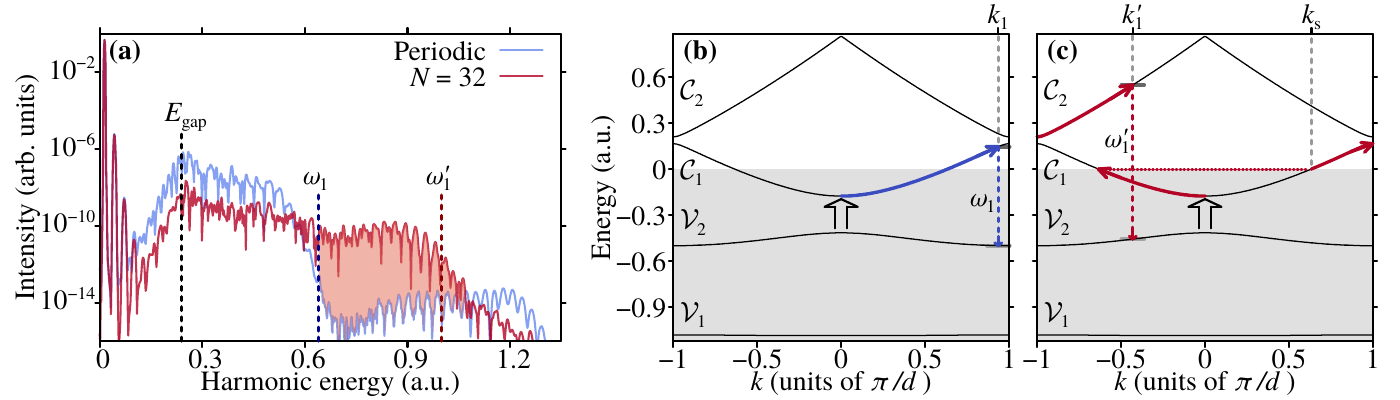}
\caption{(a) HHG spectrum $S_N$ for a finite chain of $N\,{=}\,32$ versus the spectrum in the fully periodic limit $N{\to}\infty$, driven by a laser field with parameters $A_{0} = 0.21$ and $\lambda = 3.2\,\mu$m. The vertical dashed lines indicate the first cutoff for the periodic system [$\omega_{1}$, illustrated in (b)] and the extended cutoff due to backscattering in the finite system [$\omega'_{1}$, illustrated in (c)]. The shaded area indicates the backscattering-induced enhancement of the high harmonics with energy above $\omega_{1}$. 
(b) Sketch of the $k$-space dynamics in the periodic system, for an electron excited to the bottom of {\cbone} at $A(t_{0})\,{\approx}\,{-}A_{0}$. The cutoff $\omega_{1}$ is given by the maximal {\cbone--\vbtwo} band energy difference achieved at $k_{1}\,{=}\,{2A_{0}}$. 
(c) Sketch of the $k$-space dynamics in the finite system, with an edge backscattering event in {\cbone} at the vacuum level occurring at $A(t_{\mathrm s})\,{=}\,{-}A_{0}$. The horizontal dotted line represents the sign change of $k(t)$ due to backscattering. With a subsequent band-gap transition to {\cbtwo}, this sketch corresponds to the maximally achievable harmonic energy in the backscattering case $\omega'_{1}$ at $k'_{1}=k_{\mathrm s}{+}2A_{0}{-}2\pi/d$ with $\cbone(k_{\mathrm s})=0$.
}
\label{fig:recap_backscat}
\end{figure*}%

\section{Backscattering effect on HHG in a finite chain}\label{sec:recap}
\noindent
For our present context, we briefly state how edge backscattering, as introduced recently \cite{yusa+21}, extends the HHG cutoff and enhances the high harmonics generated by delocalized electrons.
To this end, we present a comparison of the HHG spectra for the fully periodic case $N{\to}\infty$ and for the backscattering case represented by a finite chain of $N\,{=}\,32$, as shown in Fig.\,\ref{fig:recap_backscat}a.
The laser parameters, cf.\ Eq.\,\eqref{eq:vecpot}, considered in this example are $A_{0}\,{=}\,0.21$, $\omega_{0}\,{=}\,0.01425$ (corresponding to a wavelength of $\lambda\,{=}\,3.2\,\mu$m), and $n_\text{cyc}\,{=}\,9$.

Both spectra in Fig.\,\ref{fig:recap_backscat}a exhibit a peak around the energy corresponding to the {\cbone--\vbtwo} band gap,
$E_{\text{gap}} \equiv \cbone(k\,{=}\,0)-\vbtwo(k\,{=}\,0) = 0.24$,
indicating that the harmonics above this energy gap are dominated by interband processes. 
The spectrum in the fully-periodic limit manifests two plateaus: the 1st plateau, which is the primary one, has its spectral intensity significantly higher than the 2nd one by ${\sim}6$ orders of magnitude.
The 1st cutoff for the periodic system can be estimated as the maximal {\cbone--\vbtwo} band energy difference $\omega_{1}=\cbone(k_{1})-\vbtwo(k_{1}) = 0.64$ with $k_{1}\,{=}\,2A_{0}$ the largest momentum gain through unperturbed interaction with the laser field, as illustrated in Fig.\,\ref{fig:recap_backscat}b.
Note that the lowest valence band {\vbone} does not participate actively in the HHG processes due to the large gaps to other bands.

Turning to the finite chain of $N\,{=}\,32$, the spectrum shows an extended plateau, and the high harmonics with energy above $\omega_{1}$ get noticeably enhanced, see the shaded area in Fig.\,\ref{fig:recap_backscat}a.
Our previous study \cite{yusa+21} has revealed that such an effect is due to backscattering from the edge of the chain, and it typically occurs when the chain length is comparable to the full quiver excursion of the excited electron.
Backscattering allows the electron to acquire a larger momentum through the sign change of $k(t)$ and opens a pathway to high-energy states (e.g., in the 2nd conduction band {\cbtwo}).
The band energy at backscattering should be below the vacuum level to avoid ionization \cite{yusa+21}, which defines the maximal momentum at backscattering $k_{s}\,{=}\,0.285$ fulfilling $\cbone(k_{s})=0$.
Therefore, the maximally achievable harmonic energy due to backscattering is $\omega'_{1}=\cbtwo(k'_{1})-\vbtwo(k'_{1}) = 1.0$ with $k'_{1}\,{=}\,k_{s}{+}2A_{0}{-}2\pi/d$, as illustrated in Fig.\,\ref{fig:recap_backscat}c.
Note that the band picture has allowed us to provide an intuitive understanding of the HHG spectral features \cite{yusa+21}, albeit the band structure is only approximate for finite chains. 

The edge of a finite system is just a particular example of broken translational symmetry. In this work, we will explore whether enhancement of the high harmonics can be achieved by other means of breaking the periodicity.
For this purpose we use the double-chain introduced in Sec.\,\ref{sec:theory}, which is simple but insightful in capturing different types of periodicity breaks.
By varying $\delta$, the gap between the two subchains, we will systematically investigate how this internal boundary influences HHG.

\begin{figure*}[t]
\includegraphics[width=\textwidth]{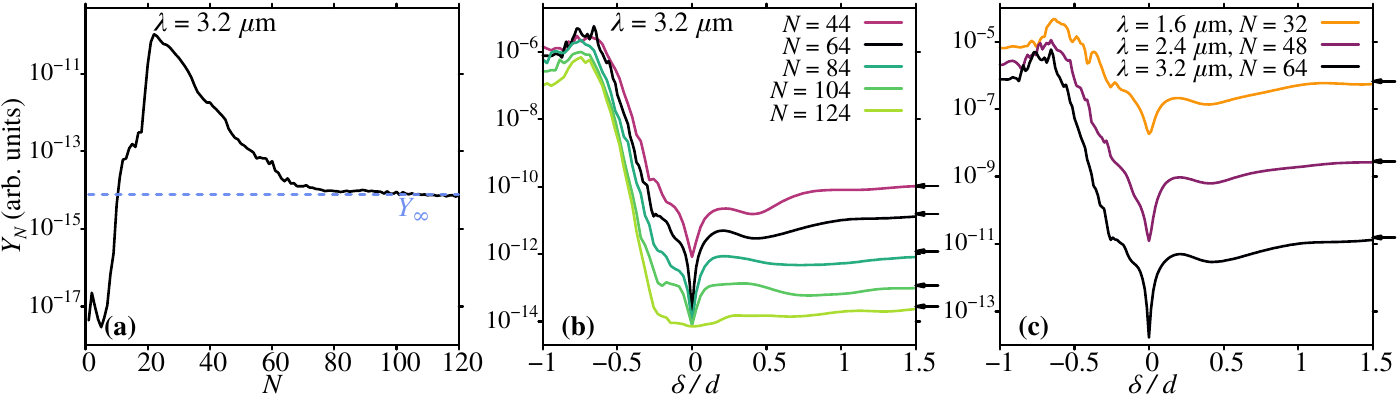}
\caption{(a) The integrated yield $Y_{N}$ beyond the periodic system HHG cutoff $\omega_{1}$, as a function of $N$ for single chains (i.e., in the special case of $\delta=0$) driven by a laser field with wavelength $\lambda=3.2\,\mu$m.
The horizontal dashed line indicates the value in the fully-periodic limit $N{\to}\infty$.
(b,\,c) $Y_{N}$ as a function of the gap $\delta$ in double-chains. 
The curves in (b) show the results for different $N$ at a fixed wavelength $\lambda=3.2\,\mu$m;
while those in (c) show the results for three different wavelengths with a fixed ratio of $N/\lambda$.
All the calculations are done with a fixed vector potential amplitude $A_{0}=0.21$.
The arrows on the right of panels (b) and (c) indicate the results obtained from isolated single chains with $N/2$ atoms, which correspond to the limit of $\delta{\to}\infty$.
}\label{fig:yield_all}
\end{figure*}%

\section{HHG in double-chains}\label{sec:result}
\subsection{Integrated yield of the high harmonics}\label{subsec:yield}
\noindent
As detailed above, we are interested in the enhancement of high harmonics with energies above the 1st cutoff $\omega_{1}$ for the periodic system.
To quantify the enhancement, we integrate the harmonic yield beyond $\omega_{1}$, namely
\begin{equation}\label{eq:yield_N}
Y_{N}=\tint_{\!\!\!\omega_{1}}\!\d{\omega}\ S_{N}(\omega).
\end{equation}
Figure~\ref{fig:yield_all}a illustrates $Y_{N}$ for single chains at laser parameters we also use in the following, where we will see that the change of $Y_{N}$ induced by the gap $\delta$ in a periodicity-broken chain of $N$ atoms is a convenient indicator of possible HHG enhancement. 

In order to investigate the dependence of $Y_{N}$ on $\delta$ a choice of $N$ promises most insight,
where the separated chains with $N/2$ atoms for large $\delta$ exhibit strong backscattering enhanced HHG, while the double-chain of length $N$ at vanishing gap $\delta\,{=}\,0$ is too long to do so. We see from Fig.\,\ref{fig:yield_all}a that pronounced enhancement (e.g., where $Y_{N}$ is higher than $Y_{\infty}$ by more than one order of magnitude) occurs for single chains containing between ${\sim}15$ and ${\sim}60$ atoms with the maximum around 20 atoms. Therefore we have chosen double-chains with a length roughly between 40 and 120 atoms in Fig.\,\ref{fig:yield_all}b.
Since for the yield $Y_{64}$ the double-chain with vanishing gap $\delta\,{=}\,0$ is almost too long for backscattering, while the separated chains with 32 atoms are close to maximal enhancement, we see here the deepest dip. The value of $Y_{44}$ at the dip nearly coincides with the value of $Y_{84}$ at large $\delta$, since there separated chains have with length 42 almost the same number of atoms contributing to backscattering-type HHG. The lowest curve $Y_{124}$ is basically flat for $\delta>0$ since even the separated chains with 62 atoms are too large for significant backscattering, much more so the gapless chain with 124 atoms.
We note that at a separation of $\delta\,{=}\,1.5d$ the limit of separated chains is almost reached, whose HHG yield are indicated with arrows and can be read off Fig.\,\ref{fig:yield_all}a.

Turning now to negative $\delta$, we see a radically different behavior with a large enhancement of $Y_{N}$ following a uniform pattern with a slope almost independent of $N$.
This suggests a mechanism for enhancement different from backscattering. To underline the difference, we present in Fig.\,\ref{fig:yield_all}c a scenario where the conditions for backscattering are the same, since a fixed ratio $N/\lambda$ is chosen. Now, indeed the shapes of the curves in the backscattering domain $\delta\,{>}\,0$ are almost the same, while the slopes are quite different for negative $\delta$. In this scenario, the tunneling probability to the conduction band is quite different since at fixed vector potential $A_{0}$ a smaller wavelength implies larger field strength and therefore larger tunneling rates in agreement with the ordering of the three yield curves.

We finally note, that a close look on Fig.\,\ref{fig:yield_all}c reveals a transition region $-0.3d\,{\lesssim}\,\delta\,{<}\,0$ whose character will become clearer in the following.

\subsection{HHG spectra}\label{subsec:spectra}
\begin{figure}[tb]
\includegraphics[width=\columnwidth]{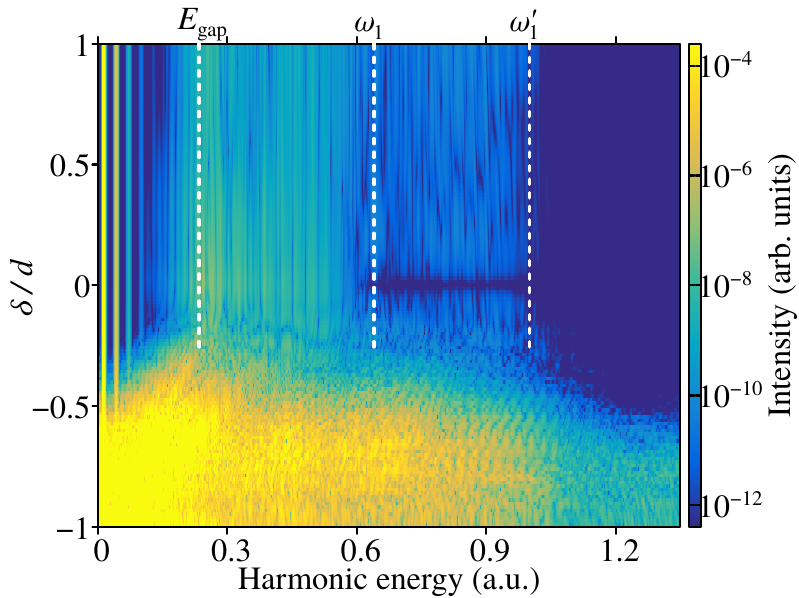}
\caption{HHG spectra for the double-chain of $N\,{=}\,64$ with the variable gap $\delta$ between subchains, for the laser wavelength of $\lambda = 3.2\,\mu$m and vector potential amplitude $A_{0} = 0.21$.
This plot corresponds to the curve with the darkest color in Figs.\,\ref{fig:yield_all}b and \ref{fig:yield_all}c.
}\label{fig:hhgs_map_all}
\end{figure}

\noindent
To understand better how the gap $\delta$ in the double-chain causes the enhancement of the high-harmonic yield $Y_{N}$ we present in Fig.\,\ref{fig:hhgs_map_all} the variation of HHG spectra with $\delta$ for fixed $N\,{=}\,64$ and $\lambda = 3.2\,\mu$m.
It clearly reveals two qualitatively different regions: Region I, covering all positive $\delta$ and a smaller range of negative $\delta$, is structured by characteristic energies (white dashed lines) independently of the value of $\delta$, namely the energy of the bandgap, $E_{\rm gap}$, and the standard ($\omega_{1}$) as well as the backscattering enhanced ($\omega'_{1}$) cutoffs. Region II, adjacent to the first one covering the remaining range of negative $\delta$ exhibits a strong, structureless enhancement of high harmonics without a clear cutoff, already familiar from Figs.\,\ref{fig:yield_all}b and \ref{fig:yield_all}c. The transition between both regions happens roughly near $\delta\,{=}\,{-}0.3d$ but varies depending on the harmonic energy.
Interestingly, region I, where backscattering with its characteristic
enhanced cutoff $\omega'_{1}$ is operative, extends into negative $\delta$ which could not be clearly deduced from Figs.\,\ref{fig:yield_all}b and \ref{fig:yield_all}c.
Very obvious from Fig.\,\ref{fig:hhgs_map_all} through the minimum in the HHG yield at $\delta\,{=}\,0$ for energies beyond $\omega_{1}$ is also, that the gapless chain
is too long to exhibit backscattering harmonics and therefore its HHG spectrum has the cutoff $\omega_{1}$ of the periodic system.

Since the spectral intensity varies significantly in region II, it is difficult to identify possible cutoff energies directly from Fig.\,\ref{fig:hhgs_map_all}.
Yet, after uncovering the HHG mechanism in that regime, we will be able to identify the corresponding characteristic harmonic energies, see Sec.\,\ref{subsec:tunnel_rate} below. 

\subsection{Mechanisms of HHG in double-chains}\label{subsec:mechanisms}
\noindent
The presence of the gap $\delta$ modifies the electronic properties of the double-chain, such as the KS potential and orbitals. Details are given in the Appendix.
For HHG relevant is the emergence of a vacuum-like region near the gap between subchains which increases for increasing $\delta$ [Figs.\,\ref{fig:pot_gap_n48}a and \ref{fig:pot_gap_n48}b]. It can reflect electrons similarly to an edge of an isolated single chain. Towards more negative $\delta$, the highest occupied orbital becomes an impurity state, energetically isolated from the valence band and closer to the conduction band [Fig.\,\ref{fig:pot_gap_n48}c].
With these facts in mind, in the following we will identify the HHG mechanisms from several representative HHG spectra.

\subsubsection{The backscattering mechanism}
\begin{figure}[tb]
\includegraphics[width=\columnwidth]{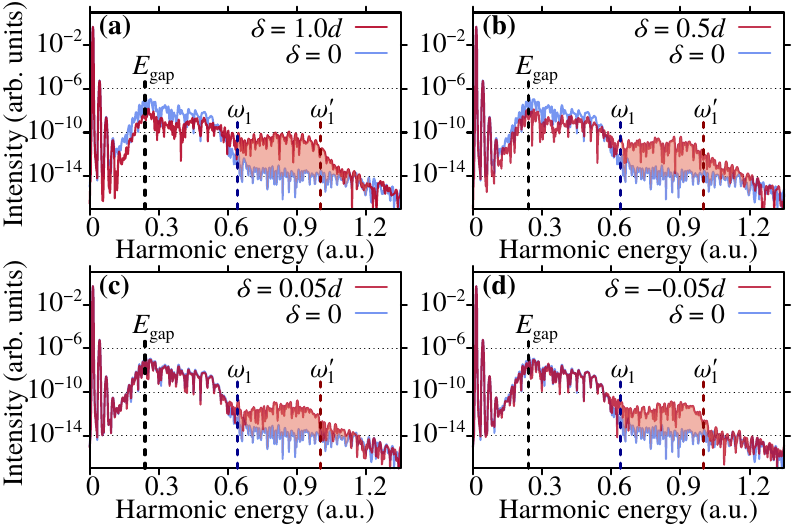}
\caption{HHG spectra for some nonzero $\delta$ values selected from Fig.\,\ref{fig:hhgs_map_all} ($N{=}64$, $\lambda{=}3.2\mu$m), compared with the spectrum for $\delta\,{=}\,0$.
The shaded area indicates the backscattering-induced enhancement of the high harmonics with energy above $\omega_{1}$.}
\label{fig:hhg_backscat}
\end{figure}

\noindent
As already inferred,
the HHG spectra in region I, $\delta\,{\gtrsim}\,{-}0.3d$, $\delta\,{\ne}\,0$, can be attributed to the backscattering mechanism.
This is confirmed in the traditional representation of Fig.\,\ref{fig:hhg_backscat}, where we contrast some selected spectra in region I with the one for $\delta\,{=}\,0$ which is taken as the reference with absent backscattering.

The spectrum for the double-chain of $N\,{=}\,64$ with $\delta\,{=}\,d$ shown in Fig.\,\ref{fig:hhg_backscat}a is very similar to that for a single chain of $N\,{=}\,32$ shown in Fig.\,\ref{fig:recap_backscat}a.
This can be understood from the effect of $\delta$ on the KS potential: As shown in Figs.\,\ref{fig:pot_gap_n48}a and \ref{fig:pot_gap_n48}b in the Appendix,
a vacuum-like region between the two subchains can effect backscattering of electrons similarly as an edge of an isolated single chain.

Note that the double-chain configuration with $\delta\,{=}\,d$ is equivalent to introducing a vacancy in the center of a single chain with $N{+}1$ atoms.
Therefore, periodicity breaks in form of vacancies offer the possibility for the backscattering mechanism and therefore extended HHG plateaus. The mechanism seems robust against possible lattice relaxations around the vacancies since it does not require an exact fulfillment of $\delta\,{=}\,d$.
For $\delta\,{=}\,0.5d$, the vacuum-like region between subchains is not yet fully developed, see Fig.\,\ref{fig:pot_gap_n48}b in the Appendix.
The corresponding HHG spectrum in Fig.\,\ref{fig:hhg_backscat}b manifests the extended plateau characterized by the same (backscattering-type) cutoff $\omega'_{1}$, with the high-harmonic signals slightly weaker than in the case of $\delta\,{=}\,d$.

Interestingly, the extended plateau due to backscattering can be observed even for very small nonzero $\delta$: 
The situation of $\delta\,{=}\,{\pm}0.05d$ shown in Figs.\,\ref{fig:hhg_backscat}c and \ref{fig:hhg_backscat}d can be considered as a weak perturbation to the regular chain, which is consistent with the fact that the harmonic signals in the primary plateau part (up to $\omega_{1}$) only differ marginally from the $\delta\,{=}\,0$ case.
The observation of the extended plateau (from $\omega_{1}$ to $\omega'_{1}$) for $\delta\,{=}\,{\pm}0.05d$ indicates that even a weak perturbation to the translational symmetry can contribute to the backscattering-type HHG, although the effect is less pronounced than that for vacuum-like internal boundaries (e.g., $\delta\gtrsim d$).

\subsubsection{The impurity-state mechanism}
\noindent
Gaps with $\delta\,{<}\,0$ give rise to impurity states that are energetically isolated from the bands (see Appendix).
In particular, the highest-occupied orbital, referred to as HOMO in the following discussion, is an impurity state with energy $\varepsilon_{\text{HOMO}}$ between the valence band {\vbtwo} and the conduction band {\cbone} as shown in Fig.\,\ref{fig:pot_gap_n48}c.
Since this (new) gap between the conduction band and the HOMO energy level, 
$E'_{\text{gap}}(\delta) \equiv \cbone(k\,{=}\,0)-\varepsilon_{\text{HOMO}}(\delta)$,
is smaller than the {\cbone--\vbtwo} band gap $E_{\text{gap}}$ defined in Sec.\,\ref{sec:recap},
the HOMO has a larger probability to tunnel to the conduction bands and can therefore significantly contribute to HHG.
Note that the impurity-driven HHG originates from the impurity state which is localized in real space, cf.\ Fig.\,\ref{fig:ee_wf_n48}b, in contrast to interband HHG (of both normal- and backscattering-type) driven by delocalized transitions.

In fact, the overall enhancement of the HHG in region II, $-d\,{\leq}\,\delta\,{\lesssim}\,{-}0.3d$, 
is similar to the effect of donor-type doping reported previously \cite{yuha+19} with a fixed HOMO energy.
The double-chain considered here offers through the continuous variation of the HOMO energy with the single parameter $\delta$ insight how the impurity-state mechanism emerges in the total HHG response.
To uncover the role of the HOMO, we separate the total current Eq.\,\eqref{eq:cur_tot} into the contribution of the highest occupied impurity orbital ($j=N_\text{occ}$) and that of all the other orbitals ($j=1,\cdots,N_\text{occ}{-}1$),
and then calculate the corresponding HHG spectra respectively.
Note that the two contributions should be added coherently in calculating the total HHG spectrum.
Yet, if for the high harmonics of interest, the spectral intensity obtained from one contribution is much higher than that obtained from the other one,
we can conclude that the former is dominant.
Such a procedure can highlight the importance of the highest occupied impurity orbital in the donor-doped case \cite[Fig.\,4]{yuha+19}.

\begin{figure}[tb]
\includegraphics[width=\columnwidth]{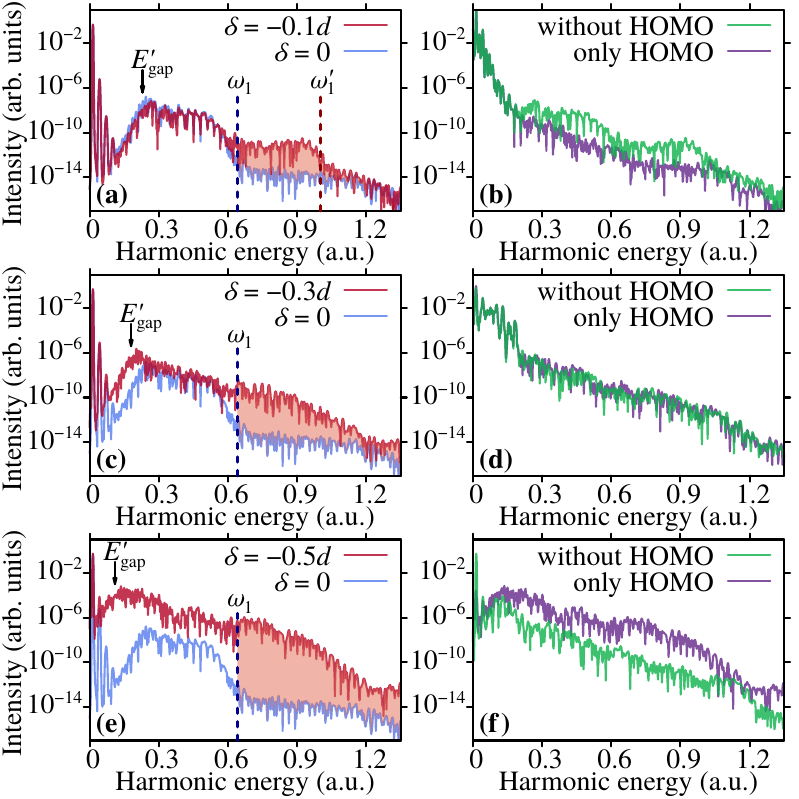}
\caption{(a,\,c,\,e) HHG spectra selected from Fig.\,\ref{fig:hhgs_map_all} at $\delta = -0.1d$, $-0.3d$, $-0.5d$, compared with the spectrum for $\delta\,{=}\,0$.
The shaded area indicates the enhancement of the high harmonics with energy above $\omega_{1}$.
The arrows with label $E'_{\text{gap}}$ indicate the gap between the {\cbone} band and the highest occupied impurity orbital.
(b,\,d,\,f) The HHG spectra calculated from only the highest occupied orbital (only HOMO) and from all the other orbitals (without HOMO), for the negative $\delta$ values in (a,\,c,\,e).}
\label{fig:hhg_impurity}
\end{figure}%

Depending on the significance of the HOMO contribution to the high harmonics, we can classify the HHG behavior into three scenarios,
visualized in Fig.\,\ref{fig:hhg_impurity} for $\delta = -0.1d$, $-0.3d$, $-0.5d$, respectively.\\
(i) In the case of $\delta\,{=}\,{-}0.1d$, as one can see in Figs.\,\ref{fig:hhg_impurity}a and \ref{fig:hhg_impurity}b,
the enhancement of the high harmonics with energy above $\omega_{1}$ is well characterized by the backscattering-type HHG cutoff $\omega'_{1}$,
and can be approximately described without the HOMO contribution.
Therefore the backscattering mechanism dominates the total HHG response,
while the impurity-state mechanism plays a negligible role in such a scenario.\\
(ii) For the intermediate case of $\delta\,{=}\,{-}0.3d$ shown in Figs.\,\ref{fig:hhg_impurity}c and \ref{fig:hhg_impurity}d,
we find that the HOMO contribution is comparable to that of other orbitals.
This means that both, the backscattering mechanism and the impurity-state mechanism, are important for the total HHG.\\
(iii) In the case of $\delta\,{=}\,{-}0.5d$, Figs.\,\ref{fig:hhg_impurity}e and \ref{fig:hhg_impurity}f clearly demonstrate the overall enhancement of the HHG, which mainly stems from the HOMO contribution.
Hence, the impurity-state mechanism dominates the total HHG response in this case.

The difference between the cases (i) and (iii) lies in the $\delta$-dependent energy of the HOMO [see, e.g., Fig.\,\ref{fig:pot_gap_n48}c in the Appendix].
The impurity-state energy $\varepsilon_{\text{HOMO}}$ for $\delta = -0.1d$ is only slightly higher than the top of valence band, 
while $\varepsilon_{\text{HOMO}}$ for $\delta = -0.5d$ is about in the middle of the band gap.
This is also reflected in the shift of the spectral peak around $E'_{\text{gap}}$ as observed through comparison of Figs.\,\ref{fig:hhg_impurity}a, \ref{fig:hhg_impurity}c and \ref{fig:hhg_impurity}e.

It is well known that the tunneling probability, which is crucial for the HHG, is exponentially sensitive to the energy gap \cite{ke64}.
Hence we expect the behavior of the HHG spectra as well as the yield $Y_{N}$ for the impurity-state mechanism to be rationalized with the $\delta$-induced change in the tunneling rate. 
This will be further analyzed below. 

\subsection{Quantitative analysis of the impurity-state mechanism}\label{subsec:tunnel_rate}
\begin{figure}[tb]
\includegraphics[width=\columnwidth]{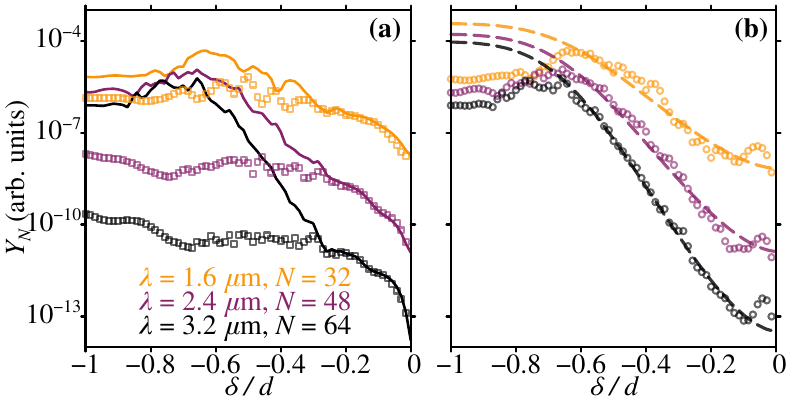}
\caption{(a) Solid lines: $Y_{N}$ calculated from the full HHG spectra (including the contributions of all the occupied orbitals),
which are the same as the results in the region of $\delta\,{<}\,0$ shown in Fig.\,\ref{fig:yield_all}c.
Square markers: $Y_{N}$ calculated with the HOMO contribution excluded from the HHG spectra.
(b) Circle markers: $Y_{N}$ calculated with only the HOMO contribution included in the HHG spectra.
Dashed lines: the tunneling exponential factor Eq.\,\eqref{eq:keldysh_rate_exp}.
}
\label{fig:yield_impurity}
\end{figure}

\noindent
Having identified the impurity-state mechanism for $\delta\,{<}\,0$ with the HHG spectra from the partial current, i.e., without the HOMO and with only the HOMO contribution,
now we apply this procedure to calculate the corresponding $Y_{N}$ distinguishing both contributions.
For the $\delta\,{<}\,0$ cases at the three different wavelengths $\lambda$ presented in Fig.\,\ref{fig:yield_all}c,
we make a comparison between the $Y_{N}$ curves computed with all the orbitals and without the HOMO contribution in Fig.\,\ref{fig:yield_impurity}a.
The obvious discrepancy in the range of $-d\,{\leq}\,\delta\,{\lesssim}\,{-}0.3d$ confirms that the HOMO contribution stemming from the impurity-state mechanism is essential for the total HHG response in this regime.

To comprehend the trend of $Y_{N}$ curves in the regime of the impurity-state mechanism,
we attempt to link the change in $Y_{N}$ to an analytical estimate of the tunneling rate.
According to the tunneling theory formulated by Keldysh \cite{ke64},
we consider an exponential factor for the overall HHG intensity given by the tunneling rate,
\begin{equation}\label{eq:keldysh_rate_exp}
R\propto\displaystyle \exp\bigg({-}\tfrac{\pi}{2}\tfrac{\sqrt{m_{*}\Delta_{*}^{3}}}{F_{0}}\Big(1-\tfrac{\omega_{0}^{2}m_{*}\Delta_{*}}{8F_{0}^2}\Big)\bigg),
\end{equation}
where $F_{0} = A_{0}\omega_{0}$ is the peak field strength the laser, $m_{*}$ is the reduced effective mass of the electron and hole, and $\Delta_{*}$ denotes the band gap.
Note that such a form was derived for direct valence- to conduction-band transitions \cite{ke64};
here we simply assume that it applies also for tunneling from the impurity state to the conduction band, by treating the impurity level as a flat band with an infinite effective mass for the hole.
Hence for our scenario, $m_{*}$ is the effective mass of the electron at the bottom of the {\cbone} band which takes the value 0.167, and $\Delta_{*} = E'_{\text{gap}}(\delta)$, the energy gap between the conduction band and the impurity level, which depends on $\delta$ monotonically (see Fig.\,\ref{fig:pot_gap_n48}c in the Appendix).

In Fig.\,\ref{fig:yield_impurity}b, the $\delta$ dependence of the tunneling exponential factor $R$, is compared with $Y_{N}$ calculated with only the HOMO contribution.
Except for $-d{\leq}\,\delta\,{\lesssim}\,{-}0.7d$, where $Y_{N}$ gets slightly attenuated when further narrowing the energy gap $E'_{\text{gap}}(\delta)$, 
we find that the tunneling exponential factor $R(\delta)$ can roughly capture the general slopes of $Y_{N}$ calculated with only the HOMO contribution.
Note that the $\delta$ dependence of the energy gap $\Delta_{*} = E'_{\text{gap}}(\delta)$ is almost identical for the different $N$ values considered here.
Therefore the difference in the slopes of the $R(\delta)$ curves is due to the change of $F_{0}$ when varying the wavelength with $A_{0}$ fixed.

In fact, the region $-d{\leq}\,\delta\,{\lesssim}\,{-}0.7d$ where the trend of $Y_{N}$ cannot be described solely with the tunneling factor $R(\delta)$, corresponds to very strong excitation which leads to depletion of the HOMO.
This depletion effect is apparent from the population of the HOMO at the end of the laser pulse.
For example, the remaining population of the HOMO in the case of $\lambda = 3.2\,\mu$m is ${\sim}95\%$ for $\delta\,{=}\,{-}0.6d$, ${\sim}55\%$ for $\delta\,{=}\,{-}0.7d$, and further drops to ${\sim}5\%$ for $\delta\,{=}\,{-}0.8d$.
Depletion of the initial state is known to pose a limitation for the HHG efficiency in atoms and molecules,
and this is also true for the impurity-state HHG in our scenario.

\begin{figure}[h]
\includegraphics[width=\columnwidth]{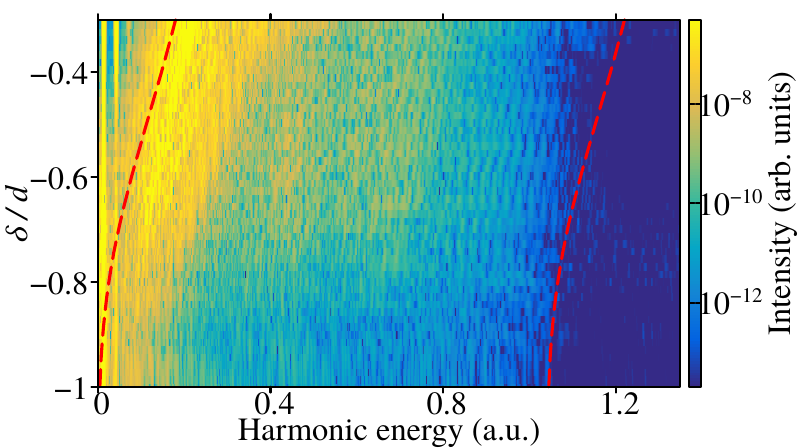}
\caption{HHG spectra with intensity scaled according to the tunneling rate for $-d\,{\leq}\,\delta\,{\leq}\,{-}0.3d$; see text for detail. The dashed lines indicate the minimal energy difference between {\cbone} and the highest occupied impurity state, and the maximal energy difference between {\cbtwo} and the highest occupied impurity state, respectively.}
\label{fig:hhgs_map_impurity}
\end{figure}

Lastly, we identify the characteristic energies for HHG spectra in the region of $-d\,{\leq}\,\delta\,{\leq}\,{-}0.3d$ with the help of the tunneling exponential factor.
As demonstrated above, the HHG spectra in this region are dominated by the HOMO contribution, and the $\delta$-induced change in the HHG signals is to a large extent captured by $R(\delta)$.
Hence we multiply the HHG spectrum for each $\delta$ in the region of $-d\,{\leq}\,\delta\,{\leq}\,{-}0.3d$ by a factor of $R(-0.3d)/R(\delta)$, to scale the spectral intensity to a similar level.
As $\delta$ varies from $-0.3d$ to $-d$, the energy gap $E'_{\text{gap}}$ decreases, and we expect that the impurity-state HHG cutoff follows a similar trend of shifting to lower energy.
Indeed, one can see from the intensity-scaled spectra that, the impurity-induced overall enhancement has a ``tail'' which basically follows the maximal energy difference between {\cbtwo} and the impurity level, $E'_{\text{max}}(\delta) \equiv \cbtwo(k\,{=}\,0)-\varepsilon_{\text{HOMO}}(\delta)$, shown as the right dashed line in Fig.\,\ref{fig:hhgs_map_impurity}.
Also, the shifting trend of the spectral peak around the energy gap $E'_{\text{gap}}(\delta)$, indicated by the left dashed line in Fig.\,\ref{fig:hhgs_map_impurity}, is better visualized than in Fig.\,\ref{fig:hhgs_map_all}.

\section{Conclusion}\label{sec:conclu}
\noindent
Solving the many-electron dynamics in double-chains based on DFT,
we have demonstrated enhancement of high-order harmonics due to periodicity breaks in the system.
Our double-chain model, with a variable separation $\delta$ between the subchains, offers a unified framework for studying the influence of broken translational symmetry in different scenarios:
For example, a negative $\delta$ gives rise to occupied impurity states similarly to the donor-type doping case, while a particular choice of setting $\delta$ equal to the lattice constant implements the situation of vacancy defects.

We have identified two mechanisms responsible for the HHG enhancement.
One is backscattering of delocalized electrons: the internal boundary caused by nonzero $\delta$ values divide the delocalized electrons into subsystems, and plays the role of reflecting the electrons similarly to the system edge.
The backscattering-type HHG (featured by an extended cutoff) exists even for tiny $|\delta|$ values characterizing a weak perturbation to the translational symmetry, and this enhancement effect builds up significantly as $\delta$ increases to positive values close to the lattice constant.
This implies that vacancy-type internal boundaries, with subsystem sizes suitable for the backscattering mechanism, can boost the efficiency of high harmonics.

The other mechanism is HHG originating from an impurity state, which is localized in real space and has an isolated energy between the valence and conduction bands.
If the impurity-state mechanism dominates the total HHG response, one typically sees an overall HHG enhancement due to the significant increase of the tunneling excitation into the conduction band.
We have found that when the gap between the conduction band and the impurity state becomes smaller,
the HHG enhancement can be roughly characterized by a tunneling exponential factor until it gets attenuated by the depletion effect.

This study confirms and extends our previous finding that HHG in a solid-like environment is sensitive to suitably broken translational symmetry. This opens a promising perspective in the rapidly expanding research area of solid-state HHG.

\appendix*
\section{The KS potential and field-free electron states of double-chains}
\noindent
In order to facilitate the understanding of the HHG behavior discussed in Sec.\,\ref{sec:result}, this appendix illustrates the $\delta$-induced changes in the field-free properties of double-chains.
Most of the observations below, such as the $\delta$-induced internal boundary in the KS potential around $x\,{=}\,0$,
the impurity-state energies for $\delta\,{<}\,0$, and the general trend of changes in the bulk-state energy levels,
are insensitive to the different system sizes $N$ considered in this work.
Basically, enlarging the number of atoms $N$ just leads to more wells in the potential and denser energy levels in the bands. 
For good visibility of these quantities, we choose a relatively small $N\,{=}\,48$ in Figs.\,\ref{fig:pot_gap_n48} and \ref{fig:ee_wf_n48} below.

\subsection{$\boldsymbol{\delta}$-induced periodicity break in the KS potential \\ and emergence of impurity states}
\begin{figure}[tb]
\includegraphics[width=\columnwidth]{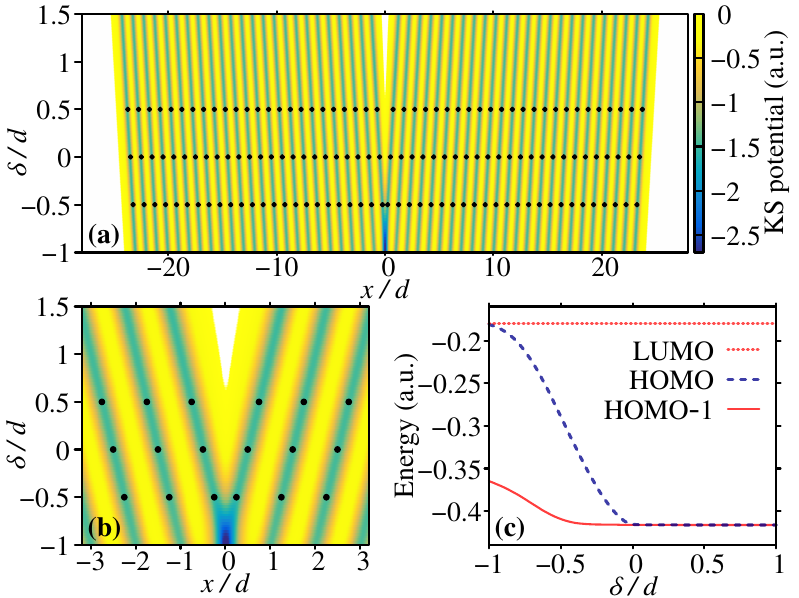}
\caption{(a) The KS potential for the double-chain of $N\,{=}\,48$ as a function of $\delta$.
The ionic positions for $\delta\,{=}\,{-}0.5d$, 0, and $0.5d$, are shown as black dots.
The white areas correspond to vacuum-like regions in which $v_\text{KS}(x)\approx 0$.
(b) Zoom-in view of (a) around the center of the system ($x\,{=}\,0$).
(c) Selected orbital energies as a function of $\delta$. The labels ``LUMO'', ``HOMO'' and ``HOMO$-$1'', as conventionally used in molecular physics, stand for the lowest unoccupied, the (1st) highest occupied and the 2nd highest occupied orbitals, respectively.}
\label{fig:pot_gap_n48}
\end{figure}

\noindent
First, we look at the static KS potential Eq.\,\eqref{eq:stat_kspot}, which determines all the orbital energies.
For a regular chain of equidistant atoms, the KS potential is periodic with the atoms located at the potential minima.
This can be seen from the special case of $\delta\,{=}\,0$ in Fig.\,\ref{fig:pot_gap_n48}a or the zoom-in view Fig.\,\ref{fig:pot_gap_n48}b.
A finite $\delta$ breaks the (quasi-)translational symmetry of the KS potential, thereby introducing an internal boundary which is of different character for positive and negative $\delta$ values, respectively.

Starting from the regular chain case at $\delta\,{=}\,0$, increasing $\delta$ implies an increasing separation of the subchains with the KS potential at $x\,{=}\,0$ approaching zero in the limit of $\delta\,{\to}\,\infty$.
The internal boundary emerges near the gap between the subchains at $x\,{=}\,0$, which gives rise to a vacuum-like region for $\delta\,{\gtrsim}\,0.7d$ shown as the white area in Fig.\,\ref{fig:pot_gap_n48}b.
The KS potential in this region is approximately zero, similarly to the vacuum region outside the system, cf.\ the white area in Fig.\,\ref{fig:pot_gap_n48}a.

For negative $\delta$, on the other hand, the two subchains are squeezed closer to each other, resulting in a deeper potential around $x\,{=}\,0$.
The two atoms located close to $x\,{=}\,0$ with indices $i\,{=}\,\tfrac{N}{2}$ and $i\,{=}\,\tfrac{N}{2}{+}1$ in Eq.\,\eqref{eq:ionloc} can be seen as an impurity introduced in the whole system which represents for the smallest value of $\delta = -d$ a doubly charged ion and therefore the deepest potential for all choices of $\delta$.

Due to the change in the KS potential, the KS orbitals are modified accordingly, manifesting different behaviors for $\delta\,{>}\,0$ and $\delta\,{<}\,0$.
For example, a positive $\delta$ hardly modifies the energy gap between unoccupied and occupied orbitals, while a negative $\delta$ gives rise to occupied impurity states energetically located within the band gap.
This is shown in Fig.\,\ref{fig:pot_gap_n48}c: 
When varying $\delta$ from 0 to $-d$, the energy of the highest occupied orbital (HOMO) becomes well separated from the HOMO$-$1 level,
and the energy gap between unoccupied and occupied orbitals gradually diminishes.
The HOMO in this case is an impurity state, which is spatially localized around the internal boundary, in contrast to bulk states which are delocalized over the entire system.

\subsection{$\boldsymbol{\delta}$-induced changes in eigen energies and wavefunctions}
\noindent
Next, we present a more detailed view of the $\delta$-induced changes in eigen energies and wavefunctions exemplified by the occupied orbitals with energy indices from $j\,{=}\,N{+}1$ to $2N$.
For a regular chain (e.g., in the special case of $\delta\,{=}\,0$), this index range $j\,{=}\,N{+}1,\cdots,2N$ includes all the bulk states in the 2nd valence band ({\vbtwo} in Fig.\,\ref{fig:recap_backscat}) and a pair of edge states with energy slightly below this band \cite{yuir+20}.
These orbitals are dominantly responsible for the high harmonics obtained from the total current Eq.\,\eqref{eq:cur_tot}, while the other lower-lying states with indices from $j\,{=}\,1$ to $j\,{=}\,N$ only make a negligible contribution to the HHG \cite{hade+17}.
The behavior revealed below remains qualitatively the same for states in other bands.

\begin{figure}[tb]
\includegraphics[width=\columnwidth]{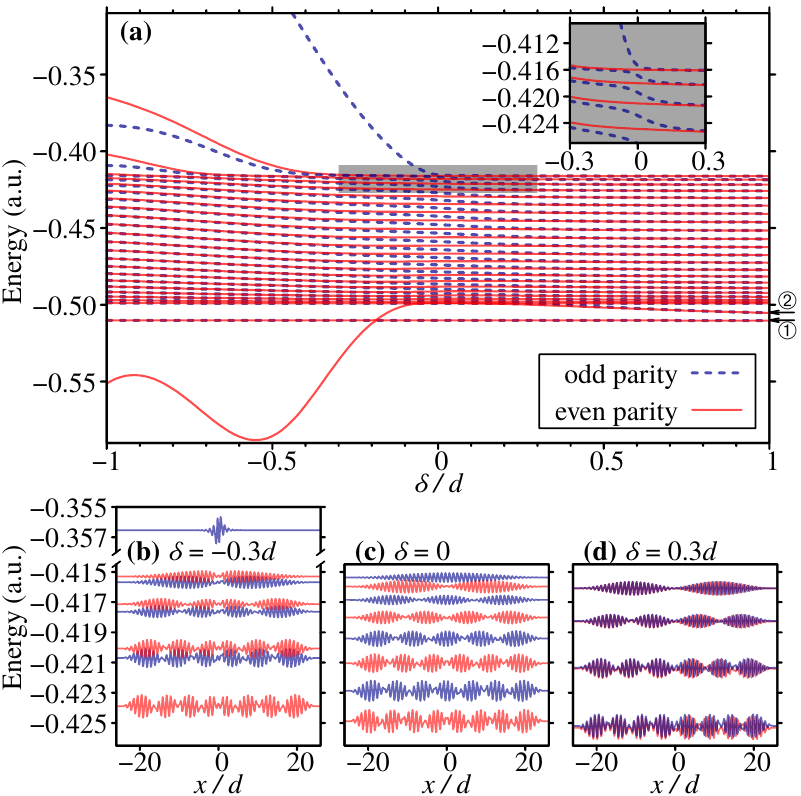}
\caption{(a) Energies of occupied orbitals with indices from $j\,{=}\,N{+}1$ to $2N$ as a function of $\delta$ for the double-chain system of $N\,{=}\,48$.
This index range includes all the states in the 2nd valence band. The inset provides a zoom-in view of the shaded area.
The arrows on the right with labels ``{\scriptsize{\textcircled 1}}'' and ``{\scriptsize{\textcircled 2}}'' indicate the ``outer'' and ``inner'' edge states, respectively; see the description in text.
(b,\,c,\,d) Wavefunctions of the eight highest occupied orbitals plotted at their corresponding energy levels, for $\delta\,{=}\,{-}0.3d$, 0, and $0.3d$, respectively. Note that the highest occupied orbital in the case of $\delta\,{=}\,{-}0.3d$ is localized in real space and has its energy isolated from the valence band.}
\label{fig:ee_wf_n48}
\end{figure}%

Figure~\ref{fig:ee_wf_n48}a shows how these orbital energies change with $\delta$;
note that the missing part of the HOMO energy level has already been shown in Fig.\,\ref{fig:pot_gap_n48}c.
The near-degenerate edge states below the {\cbtwo} band, indicated by the arrow with label {\scriptsize{\textcircled 1}},
have their energy ($-0.51$) almost unaffected by $\delta$.
Here we refer to these states as ``outer'' edge states, since they are spatially localized around the chain-end atoms with indices $i\,{=}\,1$ and $i\,{=}\,N$ in Eq.\,\eqref{eq:ionloc}.
An impurity state in the case of $\delta\,{<}\,0$, as mentioned above, is spatially localized around the internal boundary at $x\,{=}\,0$, which has a rather different spatial character in contrast to the ``outer'' edge states.
Hence, when decreasing $\delta$ from 0 to negative values, the impurity states gradually emerge from the ``in-band'' bulk states, and their location is always centered at the internal boundary ($x\,{=}\,0$).
This is the reason for the (avoided) energy-level crossing behavior near $\delta\,{\simeq}\,{-}0.2d$ at the bottom of Fig.\,\ref{fig:ee_wf_n48}a.

When increasing $\delta$ from 0 to positive values, we find that another near-degenerate states emerge with their energy approaching the ``outer'' edge states. These states for $\delta\,{>}\,0$, indicated by the arrow with label {\scriptsize{\textcircled 2}},
are spatially localized around the atoms with indices $i\,{=}\,\tfrac{N}{2}$ and $i\,{=}\,\tfrac{N}{2}{+}1$ in Eq.\,\eqref{eq:ionloc}.
Therefore we refer to them as ``inner'' edge states, only for $\delta\,{>}\,0$ where the two subchains are separated apart from each other.
Note that our double-chain model with a specific choice of $\delta\,{=}\,d$ is equivalent to a single chain with a vacancy in the center,
and the corresponding ``inner'' edge states were counted as ``defect-state orbitals'' in earlier work on vacancies \cite{irha+20}.
By varying $\delta$, the gap between the subchains in our model, now we can intuitively understand why these vacancy-defect states appear:
They essentially originate from the subchain edges near the internal boundary when separating the subchains apart, and their energy will converge to the ``outer'' edge state in the limit of $\delta{\to}\infty$.

A nonzero $\delta$ also causes changes in the bulk-state energy levels, namely near-degenerate pairs of states are formed [see, e.g., inset of Fig.\,\ref{fig:ee_wf_n48}a]. 
This trend for $\delta\,{>}\,0$ can be simply understood by considering the limit of $\delta{\to}\infty$ where the subchains are well isolated and have identical energy levels.
In the $\delta\,{<}\,0$ case, as the electron density near the internal boundary is largely contributed by the localized impurity states, effectively dividing the bulk-state electrons into subsystems, which also leads to the near-degenerate behavior of the bulk-state energy levels. 
Yet, unlike the case of $\delta\,{>}\,0$, the subsystems for $\delta\,{<}\,0$ cannot be well isolated, the corresponding near degeneracy is therefore less pronounced.

To further illustrate how the near degeneracy appears, 
we show in Figs.\,\ref{fig:ee_wf_n48}b--d the position-space wavefunctions of several highest occupied orbitals for three representative cases $\delta\,{=}\,{-}0.3d$, 0 and $0.3d$. 
For $\delta\,{=}\,{\pm}0.3d$, each pair of near-degenerate bulk states share a similar wavefunction envelope, and this envelope always shows a dip at $x\,{=}\,0$, in consistency with the existence of the internal boundary there.
In the special case of $\delta\,{=}\,0$, the near degeneracy in bulk states is completely lifted, since the two subchains perfectly merge into a single chain.
One can also observe in Fig.\,\ref{fig:ee_wf_n48}b that the highest occupied orbital in the $\delta\,{<}\,0$ case is indeed an impurity state which can be easily identified from its position-space character.

\def\articletitle#1{\emph{#1.}}

\end{document}